\documentclass[10pt,prd,showpacs,onecolumn,nofootinbib]{revtex4}
\usepackage{revsymb}
\usepackage{amsmath}
\usepackage{graphicx}

\def\3nab{\tilde{\nabla}}

\def\hsp5{\hspace{5mm}}

\def\case#1/#2{\textstyle\frac{#1}{#2}}

\def\be {\begin{equation}}
\def\ee {\end{equation}}
\def\ber {\begin{eqnarray}}
\def\eer {\end{eqnarray}}
\def\bea {\begin{eqnarray}}
\def\eea {\end{eqnarray}}

\def\bc {\begin{center}}
\def\ec {\end{center}}
\def\case#1/#2{\frac{#1}{#2}}

\newcommand{\so}{{\cal O}}
\newcommand{\sr}{{\cal R}}
\newcommand{\CR}{{\cal R}}

\newcommand{\CD}{{\cal D}}

\newcommand{\average}[1]{\left\langle #1 \right\rangle_{\CD}}

\begin{document}

\title{Spatially averaged cosmology in an arbitrary coordinate system}

\author{Julien Larena}
\affiliation{Department of Mathematics and Applied Mathematics, University of Cape Town, Rondebosch 7701, South Africa}
\email{julien.larena@gmail.com}
\date{\today}

\begin{abstract}
This paper presents a general averaging procedure for a set of observers which are tilted with respect to the cosmological matter fluid. After giving the full set of equations describing the local dynamics, we define the averaging procedure and apply it to the scalar parts of Einstein's field equations. In addition to the standard backreaction, new terms appear that account for the effect of the peculiar velocity of the matter fluid as well as the possible effect of a shift in the coordinate system. 
\end{abstract}

\maketitle

\section{Introduction}

The standard cosmological paradigm implicitly relies on a spatial averaging procedure applied to both the space-time metric and the Einstein field equations. This procedure is required to obtain an effective homogeneous model for the large scale properties of the Universe, given a clumpy local distribution of matter that is known to emerge from gravitational accretion. In the context of perturbation theory, this effective homogeneous model is referred to as the background space-time. Nevertheless, the late time Universe appears highly structured on scales up to a few hundreds of megaparsecs, but a scale of homogeneity seems to exist at larger scales \cite{HomScal}. Although this homogeneity scale arises from observations made along the past lightcone of the observer, it is usually used to support the cosmological principle, that states the existence of a homogeneity scale for the spatial distribution of matter. It is beyond the scope of this paper to address the problem of averaging over the lightcone; we will rather stick to the conservative point of view and suppose that there exists a spatial scale of homogeneity. The usual way to deal with this issue is to ignore the structures on all scales and consider the spatially locally homogeneous and isotropic solutions of Einstein field equations known as Friedmann-Lema\^itre-Robertson-Walker (FLRW hereafter) metrics. To account for the presence of structures, the model then relies on a perturbative expansion of the metric components around this FLRW background. Even though this model has proved to be extraordinarily successful in explaining a lot of features of the observed Universe, through the so-called concordance model, it suffers from a lack of theoretical basis from the point of view of the averaging it involves. Indeed, one would like to see the background emerge from the averaging procedure, rather than being postulated from the very beginning. Moreover, it is known that the nonlinearity of Einstein field equations implies that the equations for the smoothed metric are not equivalent to the smoothed equations for the actual local metric of space-time \cite{Ellis84}. In other words, the coarse-graining of the local fluctuations influences the kinematics of the effective model on large scales. This has been studied in the last decade for a simple averaging procedure, leading to the concern with the backreaction effect \cite{ThomasDust, Thomas:PerfFluid}. From a more observational point of view, this topic has recently received a revival of interest as a possible natural explanation for the Dark Energy problem (see e.g., for good reviews, \cite{Rasanen, ThomasReview}, and for recent attempts to obtain observational constraints on the homogeneous model \cite{AICpaper, Bolejko}) that tarnishes the success of the concordance model (in the sense that this latter fails to explain elegantly the origin of the phenomenon).\\
In this paper, we generalize the work done by T. Buchert about the backreaction effect for a set of observers comoving with the matter fluid \cite{Thomas:PerfFluid} by extending the averaging procedure to an arbitrary set of observers, hoping in particular that it could shed light on the gauge issue in the backreaction context (i.e. the gauge dependence of the backreaction effect in perturbation theory). Buchert formalism has been used successfully in the synchronous gauge \cite{li}, where it applies directly without ambiguity. But in the Newtonian gauge, for example, different authors disagree on the type of effect obtained \cite{Mata, Rasa,Iain}. These differences can be traced back to the fact that the different authors have used different definitions of the effective Hubble parameter. In our work, we define it with respect to the expansion of the cosmic fluid as seen by the observers. It seems that this latter quantity is more suitable to describe a physical expansion as well as to compare the gauges between each other. We aim at studying in detail the Newtonian gauge in this context in a forthcoming paper \cite{NewGaugePert}. Moreover, a recent model developed by Wiltshire \cite{Wiltshire} suggests that the choice of the set of observers relative to which the averaging is performed is a crucial feature.\\
The paper is organized as follows. Section II presents the formalism used to address the problem; we insist in particular on the foliation of space-time by a set of observers that are not comoving with the matter flow, and give the general equations governing the dynamics of both the geometry and the quantities associated with matter. In Sec. III, we will present the generalized averaging procedure that aims at keeping the main characteristics of the one introduced by T. Buchert. This will require a discussion about what really has to be average in order to define an effective Hubble parameter for the model. Thanks to this way of averaging, we will then give explicitly the set of equations that governs the kinematics of the averaged model, and comment on the new terms arising from the tilt between observers and the matter flow. Finally, we sum up the results and discuss possible and ongoing uses that can be made of this new formalism in Sec. 4.

\section{Space-time dynamics in an arbitrary coordinate system}

The cosmological model consists in a space-time ${\cal M}$ endowed with a metric tensor $g_{ab}$, and filled with a matter content characterized by its energy-momentum tensor. We will assume that gravitation is well described on all scales by general relativity, so that the metric and the matter content are linked by the Einstein field equations. The line element can be written:
\begin{equation}
ds^{2}=g_{ab}dx^{a}dx^{b}\mbox{ ,}
\end{equation}
where $(x^{a})_{a\in \{0,1,2,3\}}$ are the space-time coordinates. Throughout the paper, we will employ Latin letters of the beginning of the alphabet $(a,b,c,...,h)$ to denote space-time indices, and Latin letters of the end of the alphabet to denote spatial indices (this concept will be made clearer in the next section after the projector on spatial hypersurfaces has been introduced).

\subsection{Foliation of space-time}
Let us start by setting the notations employed to describe the 1+3 foliation of space-time that will be used in this paper. It consists of introducing a set of observers $\so (p)$ defined at each point of the space-time manifold $p\in {\cal M}$, characterized by a unit four-velocity field $n^{a}$ that is everywhere timelike and future directed, i.e. $n^{a}n_{a}=-1$. Once this four-velocity is known at each point of the manifold, one can foliate space-time by a continuous series of spacelike hypersurfaces that are simply the hypersurfaces everywhere orthogonal to the vector field $n^{a}$. Moreover, after having defined the projection tensor field on these hypersurfaces by $h_{ab}:=g_{ab}+n_{a}n_{b}$, one requires the four-velocity of observers to be irrotational, i.e. $h^{a}_{c}h^{b}_{d}n_{[a;b]}=0$; this last property ensures that the projection tensor $h_{ab}$ is also a well-defined Riemannian metric tensor for the hypersurfaces orthogonal to the four-velocity $n^{a}$.
The line element can then be written, with respect to this foliation\footnote{For a clear description of the 1+3 foliation of space-time, see, e.g., the very clear paper by Smarr \& York (in particular for a physical interpretation of the lapse and shift) \cite{Smarr} and more recently, the lecture notes \cite{EricLectures}.}:
\begin{equation}
\label{eq:ADM_metric}
ds^{2}=-(N^{2}-N_{i}N^{i})dt^{2}+2N_{i}dtdx^{i}+h_{ij}dx^{i}dx^{j}\mbox{ ,}
\end{equation}
where we have introduced respectively the lapse function $N(x^{a})$ and the shift three-vector $N^{i}(x^{a})$ such that the four-velocity of observers can be decomposed as follows:
\begin{equation}
\label{eq:rel_velADM}
n^{a}=\frac{1}{N}(1,-N^{i}) \mbox{ , } n_{a}=N(-1,0,0,0)\mbox{ .}
\end{equation}

Two other quantities characterizing the hypersurfaces orthogonal to $n^{a}$ will be very important in the following:
\begin{itemize}
\item the intrinsic curvature of the hypersurfaces: $\sr :=h^{ab}\sr_{ab}$, where $\sr_{ab}$ is the three-Ricci curvature of the hypersurfaces,
\item the extrinsic curvature (or second fundamental form): $K_{ab}:= -h^{c}_{a}h^{d}_{b}n_{c;d}$ that encodes the way the hypersurfaces are embedded in the manifold ${\cal M}$.
\end{itemize}

Up to now, we have introduced a set of observers and the purely geometrical quantities associated with it. In order to have a complete description of the cosmological space-time, one still has to prescribe the matter content of the Universe. In the following, we will assume that this matter content can be well described by a perfect fluid (not necessarily irrotational) of energy density $\rho(x^{a})$, pressure $p(x^{a})$ and 4-velocity $u^{a}(x^{b})$ (with $u_{a}u^{a}=-1$), so that its stress-energy tensor reads:
\begin{equation}
T_{ab}=(\rho +p)u_{a}u_{b}+pg_{ab}\mbox{ .}
\end{equation}
Note that in this work, the four-velocity of matter $u^{a}$ is not necessarily aligned with the four-velocity of the observers $n^{a}$, so that one can introduce a vector field $v^{a}$ corresponding to the relative velocity of the matter fluid with respect to the fundamental observers. $v^{a}$ is spacelike and orthogonal to $n^{a}$ ($v^{a}n_{a}=0$) and one has:
\begin{equation}
\label{eq:relvel}
u^{a}=\gamma (n^{a}+v^{a})\mbox{ with } \gamma =\frac{1}{\sqrt{1-v^{a}v_{a}}}\mbox{ ,}
\end{equation}
where $\gamma$ is the usual Lorentz factor (see in particular \cite{Gebbie} for a 1+3 covariant treatment of multifluids cosmology, in which the effect of the tilt between a fluid and the observers is described).

We now have all the ingredients to write down the Einstein field equations. Thanks to the 1+3 foliation, they can be separated in two different sets: the constraint equations that have to be satisfied on every hypersurface, and the evolution equations, that prescribe how the fields $(h_{ab},K_{ab})$ evolve from one hypersurface to another infinitesimally closed.
The so-called Hamiltonian constraint reads:
\begin{equation}
\label{eq:HC}
\sr -K^{i}_{j}K^{j}_{i}+K^{2}=16\pi G\epsilon+2\Lambda \mbox{ ,   }\epsilon=T_{ab}n^{a}n^{b}=\gamma^{2}\rho+(\gamma^{2}-1)p\mbox{ ,}
\end{equation}
where $K:=K^{i}_{i}$. The momentum constraint is:
\begin{equation}
\label{eq:MC}
\tilde{\nabla}_{i}K^{i}_{j}-\tilde{\nabla}_{j}K=8\pi G J_{j} \mbox{ , }J_{j}=-T_{ab}n^{a}h^{b}_{j}=\gamma^{2} (\rho+p)v_{j}\mbox{ ,}
\end{equation}
where we have defined the projected covariant three-derivative on the spatial hypersurfaces of any tensor field $t^{a...c}_{\mbox{ }\mbox{ }\mbox{ }\mbox{ }\mbox{ }d...f}$: $\tilde{\nabla}_{d}t^{a...c}_{\mbox{ }\mbox{ }\mbox{ }\mbox{ }\mbox{ }\bar{a}...\bar{c}}:=h^{e}_{d}h^{a}_{a'}...h^{c}_{c'}h_{\bar{a}}^{a''}...h_{\bar{c}}^{c''}\nabla_{e}t^{a'...c'}_{\mbox{ }\mbox{ }\mbox{ }\mbox{ }\mbox{ }a''...c''}$.
The evolution equation for the first fundamental form reads:
\begin{equation}
\label{eq:FFF}
\frac{1}{N}\partial_{t}h_{ij}=-2K_{ij}+\frac{2}{N}\tilde{\nabla}_{(j}N_{i)}\mbox{ ,}
\end{equation}
and for the second fundamental form, one has:
\begin{equation}
\label{eq:SFF}
\frac{1}{N}\partial_{t}K^{i}_{j}=\sr^{i}_{j}+KK^{i}_{j}-\Lambda\delta^{i}_{j}-\frac{1}{N}\tilde{\nabla}_{j}\tilde{\nabla}^{i}N+\frac{1}{N}\left(K^{i}_{k}\tilde{\nabla}_{j}N^{k}-K_{j}^{k}\tilde{\nabla}_{k}N^{i}+N^{k}\tilde{\nabla}_{k}K^{i}_{j}\right)-8\pi G \left(S^{i}_{j}+\frac{1}{2}(\epsilon-S^{k}_{k})\delta^{i}_{j}\right)\mbox{ ,}
\end{equation}
with $S_{ij}=\gamma^{2}\rho v_{i}v_{j}+p(h_{ij}+\gamma^{2}v_{i}v_{j})$.
These equations have to be supplemented by the energy-momentum conservation for the matter fluid: $\nabla_{a}T^{a}_{b}=0$.
We will now introduce the standard decomposition for the covariant spatial derivatives of the four-vectors in terms of their trace, symmetric trace-free and antisymmetric parts. Writing $\dot{f}\equiv n^{a}\nabla_{a}f$ for any quantity $f$, one can write:
\begin{eqnarray}
\nabla_{a}n_{b}&=&-n_{a}\dot{n}_{b}+\frac{1}{3}\xi h_{ab}+\Sigma_{ab}\\
\mbox{with }& & \xi\equiv h^{c}_{a}h^{da}\nabla_{c} n_{d} \mbox{ and } \Sigma_{ab}\equiv h^{c}_{a}h^{d}_{b}\nabla_{(c}n_{d)}-\frac{1}{3}\xi h_{ab}\mbox{ ;}\nonumber\\ 
\nabla_{a}u_{b}&=&-n_{a}\left(2\dot{n}_{b}+v\dot{v}n_{b}\right)-\gamma\tilde{\nabla}_{a}v n_{b}+\frac{1}{3}\theta h_{ab}+\sigma_{ab}+\omega_{ab}\\ \mbox{with }& &\theta\equiv h^{c}_{a}h^{da}\nabla_{c} u_{d} \mbox{ , } \sigma_{ab}\equiv h^{c}_{a}h^{d}_{b}\nabla_{(c}u_{d)}-\frac{1}{3}\theta h_{ab}\mbox{ and } \omega_{ab}\equiv h^{c}_{a}h^{d}_{b}\nabla_{[c}u_{d]}\mbox{ ;}\nonumber\\
\nabla_{a}v_{b}&=&-n_{a}\left(\dot{v}_{b}+X_{b}\right)+Y_{a}n_{b}+\frac{1}{3}\kappa h_{ab}+\beta_{ab}+W_{ab}\\
\mbox{with }& & \kappa\equiv h^{c}_{a}h^{da}\nabla_{c} v_{d} \mbox{ , } \beta_{ab}\equiv h^{c}_{a}h^{d}_{b}\nabla_{(c}v_{d)}-\frac{1}{3}\kappa h_{ab}\nonumber\\
 \mbox{and }& & X_{a}\equiv n^{b}h_{a}^{c}\nabla_{c}v_{b}\mbox{ , } Y_{b}\equiv n^{c}h_{b}^{a}\nabla_{c}v_{a} \mbox{ , } W_{ab}\equiv h^{c}_{a}h^{d}_{b}\nabla_{[c}v_{d]}\nonumber\mbox{ .}
\end{eqnarray}
In these relations, $\xi$, $\theta$ and $\kappa$ denote the isotropic expansion rates of the four-velocities $n^{a}$, $u^{a}$ and of the peculiar velocity $v^{a}$ respectively, and $\Sigma_{ab}$, $\sigma_{ab}$ and $\beta_{ab}$ their shears, with respect to the threading of space-time induced by the vector $n^{a}$. $\omega_{ab}$ and $W_{ab}$ are, respectively, the vorticities\footnote{Since it defines the foliation, $n^{a}$ is vorticity free by definition.} of $u^{a}$ and $v^{a}$ in this same foliation. These quantities are those measured by the observers with four-velocities $n^{a}$ in their instantaneous rest-frame. In particular $\theta$, $\sigma_{ab}$ and $\omega_{ab}$ differ from the usual expansion, shear and vorticity of the matter fluid as measured by observers comoving with this matter fluid (by acceleration terms essentially), that are defined by the decomposition of $(g^{ac}+u^{a}u^{c})(g^{bd}+u^{b}u^{d}\nabla_{c}u_{d})$. For example, the expansions are linked by the relation:
\begin{equation}
\Theta\equiv \nabla_{a}u^{a}=\theta+\gamma (\gamma^{2}v^{a}\dot{v}_{a}-n^{a}\dot{v}_{a})\mbox{ .}
\end{equation}
Using (\ref{eq:relvel}), one can relate this quantities as follows:
\begin{eqnarray}
\xi&=&\gamma^{-1}\theta-\kappa-\gamma^2 B\\
\Sigma_{ab}&=&\gamma^{-1}\sigma_{ab}-\beta_{ab}-\gamma^2\left(B_{(ab)}-\frac{1}{3}Bh_{ab}\right)\\
W_{ab}&=&\gamma^{-1} \omega_{ab}-\gamma^{2}B_{[ab]}\mbox{ ,}
\end{eqnarray}
where we have introduced the tensor:
\begin{equation} 
B_{ab}\equiv \frac{1}{3}\kappa(v_{a}n_{b}+v_{a}v_{b})+\beta_{ca}v^{c}n_{b}+\beta_{ca}v^{c}v_{b}+W_{ca}v^{c}n_{b}+W_{ca}v^{c}v_{b}\mbox{ ,}
\end{equation}
whose trace is given by $B=\frac{1}{3}\kappa v^{2}+\beta_{ab}v^{a}v^{b}$, which encompasses the peculiar velocity effects. In our notation, angular and round brackets denote the antisymmetric and symmetric parts, respectively, of a tensor projected with $h_{a b}$.
It is important to note that the extrinsic curvature $K_{ij}$ is related to the derivative of $n^{a}$ by: $K_{ij}=-h^{a}_{i}h^{b}_{j}\nabla_{a}n_{b}$, so that the equations for the expansion and shear can be obtained by suitably replacing $K_{ij}$ in the Einstein field equations (\ref{eq:HC})-(\ref{eq:SFF}).
Let us finally introduce the following notation for convenience:
\begin{eqnarray}
\theta_{B}&\equiv& -\gamma\kappa-\gamma^{3} B\\
\sigma_{Bij}&\equiv& -\gamma\beta_{ij}-\gamma^{3}\left(B_{(ij)}-\frac{1}{3}Bh_{ij}\right)\mbox{ ,}
\end{eqnarray}
so that:
\begin{eqnarray}
\xi=\gamma^{-1}(\theta+\theta_{B})\mbox{ ,}\\
\Sigma_{ij}=\gamma^{-1}(\sigma_{ij}+\sigma_{Bij})\mbox{ .}
\end{eqnarray}

\subsection{Field equations}

With these notations in hand, we can now write down the complete system of equations describing the behavior of matter on local scales. They read:
\begin{eqnarray}
\label{eq:Eq1}
& &\gamma^{2}\CR+\frac{2}{3}\theta^{2}+\frac{2}{3}\theta_{B}^{2}+\frac{4}{3}\theta\theta_{B}-2\sigma^{2}-2\sigma_{B}^{2}=16\pi G\gamma^{4}\left(\rho+v^{2}p\right)+2\Lambda\gamma^{2}\\
\label{eq:Eq2}
& &\frac{2}{3}(\tilde{\nabla}_{j}\theta+\tilde{\nabla}_{j}\theta_{B})-\tilde{\nabla}_{i}\sigma^{i}_{j}-\tilde{\nabla}_{i}\sigma^{i}_{bj}-v\left(\frac{2}{3}(\theta+\theta_{B})\tilde{\nabla}_{j}v+(\sigma^{i}_{j}+\sigma^{i}_{bj})\tilde{\nabla}_{i}v\right)=8\pi G\gamma^{3}(\rho+p)v_{j}
\end{eqnarray}
for the constraint equations (\ref{eq:HC})-(\ref{eq:MC}). Moreover, the evolution equations deduced from (\ref{eq:FFF})-(\ref{eq:SFF}) are:
\begin{eqnarray}
\label{eq:Eq3}
& &\frac{\gamma}{N}\partial_{t}h_{ij}=\frac{2}{3}(\theta+\theta_{B})h_{ij}+2(\sigma_{ij}+\sigma_{Bij})+\frac{2\gamma}{N}\tilde{\nabla}_{(j}N_{i)}\\
\label{eq:Eq4}
& &\partial_{t}\theta=\frac{1}{\gamma}(\theta+\theta_{B})(\partial_{t}\gamma-N^{k}\tilde{\nabla}_{k}\gamma)-\partial_{t}\theta_{B}-N\gamma\CR-N\gamma^{-1}\theta^{2}-N\gamma^{-1}\theta_{B}^{2}-2N\gamma^{-1}\theta\theta_{B}+3N\gamma\Lambda\nonumber\\
& &+\gamma\tilde{\nabla}^{i}\tilde{\nabla}_{i}N+N^{k}\tilde{\nabla}_{k}(\theta+\theta_{B})+4\pi G\left(N\gamma^{3}\rho(3-v^{2})-N\gamma p(3+2\gamma^{2}v^{2})\right)\\
\label{eq:Eq5}
& &\partial_{t}\sigma^{i}_{j}=\frac{\partial_{t}\gamma-N^{k}\tilde{\nabla}_{k}\gamma}{\gamma}(\sigma^{i}_{j}+\sigma_{Bj}^{i})-\partial_{t}\sigma_{Bj}^{i}-N\gamma(\CR^{i}_{j}-\frac{1}{3}\CR\delta^{i}_{j})-\frac{N}{\gamma}(\theta+\theta_{B})(\sigma^{i}_{j}+\sigma_{Bj}^{i})\nonumber\\
& &+\gamma(\tilde{\nabla}^{i}\tilde{\nabla}_{j}N-\frac{1}{3}\tilde{\nabla}^{k}\tilde{\nabla}_{k}(N)\delta^{i}_{j})-\tilde{\nabla}_{k}N^{i}(\sigma^{k}_{j}+\sigma_{Bj}^{k})\nonumber\\
& &+\tilde{\nabla}_{j}N^{k}(\sigma^{i}_{k}+\sigma_{Bk}^{i})+N^{k}\tilde{\nabla}_{k}\sigma^{i}_{j}+N^{k}\tilde{\nabla}_{k}\sigma^{i}_{bj}+8\pi G N\gamma^{3}(\rho+p)(v^{i}v_{j}-\frac{v^{2}}{3}\delta^{i}_{j})\mbox{ ,}
\end{eqnarray}
with $\sigma^{2}=\sigma^{i}_{j}\sigma^{j}_{i}/2$ and $\sigma_{B}^{2}=\sigma^{i}_{bj}\sigma^{j}_{bi}/2+\sigma_{ij}\sigma_{B}^{ij}$.
Finally,the equations obtained from the energy-momentum conservation $\nabla_{a}T^{a}_{b}=0$ read:
\begin{eqnarray}
\label{eq:Eq6}
\frac{1}{N}\partial_{t}\rho+(v^{k}-\frac{N^{k}}{N})\tilde{\nabla}_{k}\rho+\left(\gamma^{-1}\theta +2v\dot{v}+v_{k}a^{k}\right)(\rho+p)+\frac{v^{2}}{N}\partial_{t}p+v^{2}\left(v^{i}-\frac{N^{i}}{N}\right)\tilde{\nabla}_{i}p&=&0\\
\label{eq:Eq7}
\frac{1}{N}\partial_{t}v^{k}-\frac{N^{i}}{N}\tilde{\nabla}_{i}v^{k}+\left(\frac{1}{3\gamma}\theta-v_{i}a^{i}\right)v^{k}+a^{k}+\gamma^{-1}\left(\sigma_{j}^{k}v^{j}+\omega_{j}^{k}v^{j}\right)+\frac{1}{\gamma^{2}(\rho+p)}\left(\tilde{\nabla}^{k}p+\dot{p}v^{k}\right)&=&0\mbox{ .}
\end{eqnarray}
 We have noted $a_{k}:=h_{a k}\dot{n}^{a}=(\tilde{\nabla}_{k}N)/N$ the acceleration of the observers along their world-lines.

\section{The general spatial averaging procedure}

In this paper, we will retain as our guideline the averaging formalism introduced in cosmology by T. Buchert \cite{ThomasDust, Thomas:PerfFluid}. Namely, we will consider the operator $\average{.}$ acting on scalar quantities $\psi(t,x^{i})$ such that, for any compact domain $\CD$ included in an hypersurface orthogonal to the observers worldlines:
\begin{equation}
\label{eq:averager}
\average{\psi}\equiv \frac{1}{V_{\CD}}\int_{\CD}\psi(t,x^{i})Jd^{3}x\mbox{ ,}
\end{equation}
where $J\equiv \sqrt{\det (h_{ij})}$ and $V_{\CD}\equiv \int Jd^{3}x$ is the Riemannian volume of $\CD$.
From Eq. (\ref{eq:Eq3}), one can deduce an evolution equation for the quantity $J$:
\begin{equation}
\label{eq:JEvol}
\frac{1}{J}\partial_{t}J=\gamma^{-1}N(\theta-\kappa)+\tilde{\nabla}_{k}N^{k}\mbox{ ,}
\end{equation}
that translates into a evolution equation for the volume $V_{\CD}$:
\begin{equation}
\label{eq:VolEvol}
\partial_{t}V_{\CD}=\int_{\CD}\left(\gamma^{-1}N(\theta-\kappa)+\tilde{\nabla}_{k}N^{k}\right)Jd^{3}x\mbox{ .}
\end{equation}
One can immediately see that the volume scale factor as defined in the usual Buchert framework, by $a_{\CD}^{V}=\left(V_{\CD}(t)/V_{\CD}(t_{i})\right)^{1/3}$ is no longer related in a simple way to the averaged expansion of the matter flow as measured by the observers, $\average{N\theta}$. But the measurable quantity of interest in cosmology is this latter expansion, not the scale factor, so in building an effective homogeneous cosmological model, we should concentrate on estimating the observable quantity, i.e. the expansion of matter. In the following, we will then retain as a definition for the Hubble rate of expansion:
\begin{equation}
\label{eq:HubbleDef}
3H_{\CD}\equiv \average{N\theta}=\frac{1}{V_{\CD}}\int_{\CD}N\theta Jd^{3}x\mbox{ ,}
\end{equation}
and define the effective scale factor $a_{\CD}(t)$ such that:
\begin{equation}
\label{eq:SFDef}
H_{\CD}=\frac{\partial_{t}a_{\CD}}{a_{\CD}}\mbox{ .}
\end{equation}
The relation between this effective (matter expansion) scale factor and the (observers) volume scale factor is then:
\begin{equation}
\label{eq:RelScalFac}
a_{\CD}^{V}=a_{\CD}\exp\left(\int_{ti}^{t}\left(\average{(\gamma^{-1}-1)N\theta-\gamma^{-1}N\kappa+\tilde{\nabla}_{k}N^{k}}\right)(t^{'})dt'\right)\mbox{ .}
\end{equation}
The definition (\ref{eq:SFDef}) coincides with the volume scale factor for observers comoving with the fluid (i.e. implying $\gamma=1$ and $\kappa=0$) in a coordinate system comoving with the fluid (i.e. $N^{i}=0$). This scale factor has been introduced only to rewrite the averaged equations in a form that can be compared with the usual Friedmann equations, but it is important to insist on the fact that everything could be expressed in terms of the effective Hubble expansion only, without any reference to a scale factor of any kind. It has to be noted that the observers volume scale factor $a_{\CD}^{V}$ is the one whose evolution has been studied in \cite{Iain}; it seems more natural to concentrate on the matter expansion rate as seen by observers, rather than to the expansion rate of the observers themselves.
One of the main relations in averaged cosmologies comes from the fact that averaging and evolution do not commute, which, making use of Eq. (\ref{eq:VolEvol}) translates into:
\begin{equation}
\label{eq:CommRel}
[\partial_{t}.,\average{.}]\psi(t,x^{i})=\average{\left(N\gamma^{-1}(\theta+\theta_{B})+\tilde{\nabla}_{k}N^{k}\right)\psi}-\average{N\gamma^{-1}(\theta+\theta_{B})+\tilde{\nabla}_{k}N^{k}}\average{\psi}\mbox{ ,}
\end{equation}
for any scalar function $\psi(t,x^{i})$.
One can then write the averaged equations arising from averaging the scalar parts of Einstein field equations, Eqs. (\ref{eq:Eq1}) and (\ref{eq:Eq4}):
\begin{eqnarray}
\label{eq:Averaged1}
6H_{\CD}^{2}&=&16\pi G\left(\average{\gamma^{4}N^{2}\rho}+\average{\gamma^{2}(\gamma^{2}-1)N^{2}p}\right)+2\Lambda\average{N^{2}\gamma^{2}}-\average{\gamma^{2}N^{2}\CR}-{\cal Q}_{\CD}+{\cal L}_{\CD}\\
3\frac{\partial^{2}_{t}a_{\CD}}{a_{\CD}}&=&-4\pi G\average{N^{2}\gamma\left(\gamma^{2}\rho(1+v^{2})+3p(1+2\gamma^{2}v^{2})\right)}+\Lambda\average{N^{2}\gamma}\nonumber\\
\label{eq:Averaged2}
 & &+{\cal Q}_{\CD}+{\cal P}_{\CD}+{\cal K}_{\CD}+{\cal F}_{\CD}-{\cal L}_{\CD}\mbox{ ,}
\end{eqnarray}
where we have defined the standard kinematical backreaction:
\begin{equation}
\label{eq:kinback}
{\cal Q}_{\CD}\equiv\frac{2}{3}\left(\average{(N\theta)^{2}}-\average{N\theta}^{2}\right)-2\average{N^{2}\sigma^{2}}\mbox{ ,}
\end{equation}
and additional backreaction terms as:
\begin{eqnarray}
\label{eq:otherback1}
{\cal L}_{\CD}&\equiv&2\average{N^{2}\sigma_{B}^{2}}-\frac{2}{3}\average{(N\theta_{B})^{2}}-\frac{4}{3}\average{N^{2}\theta\theta_{B}}\\
\label{eq:otherback2}
{\cal P}_{\CD}&\equiv&\average{\theta\partial_{t}N}+\average{\gamma N\tilde{\nabla}_{k}\tilde{\nabla}^{k}N}\\
{\cal K}_{\CD}&\equiv&\average{NN^{k}\tilde{\nabla}_{k}\theta}+\average{NN^{k}\tilde{\nabla}_{k}\theta_{B}}+\average{(N\gamma^{-1}\theta_{B}+\tilde{\nabla}_{k}N^{k})N\theta}-3\average{N\gamma^{-1}\theta_{B}+\tilde{\nabla}_{k}N^{k}}H_{\CD}\nonumber\\
\label{eq:otherback3}
 & &-\average{N\partial_{t}\theta_{B}}+\average{N^{2}\gamma^{-1}\dot{\gamma}\theta}+\average{N^{2}\gamma^{-1}\dot{\gamma}\theta_{B}} -2\average{N^{2}\theta_{B}^{2}}\\
{\cal F}_{\CD}&=&\frac{2}{3}\average{N^{2}\theta^{2}(\gamma^{-1}-1)}-2\average{N^{2}\sigma^{2}(\gamma^{-1}-1)}-\average{N\theta}\average{N\theta(\gamma^{-1}-1)}\nonumber\\
 & & -\frac{1}{3}\average{N^{2}\theta_{B}^{2}(\gamma^{-1}-1)}-\frac{2}{3}\average{N^{2}\theta\theta_{B}(\gamma^{-1}-1)}-2\average{N^{2}\sigma_{B}^{2}(\gamma^{-1}-1)}\mbox{ .}
\end{eqnarray}
Note that for the sake of simplicity, we have used $\dot{\gamma}\equiv n^{a}\nabla_{a}\gamma=(\partial_{t}\gamma)/N-(N^{i}\tilde{\nabla}_{i}\gamma)/N$.

Averaging the energy conservation equation (\ref{eq:Eq6}) in the same way, one can also obtain the equation for a scaled averaged energy density $N^{2}\rho$:
\begin{eqnarray}
\partial_{t}\average{N^{2}\rho}&+&3H_{\CD}\left(\average{N^{2}\rho}+\average{N^{2}p}\right)=2\average{\rho N\partial_{t}N}-[\partial_{t}.,\average{.}](N^{2}p)+\average{(N\gamma^{-1}\theta_{B}+\tilde{\nabla}_{k}N^{k})N^{2}(\rho+p)}\nonumber\\
\label{eq:Averaged3}
& &-\average{N\gamma^{-1}\theta_{B}+\tilde{\nabla}_{k}N^{k}}\left(\average{N^{2}\rho}+\average{N^{2}p}\right)-\average{N\theta (\gamma^{-1}-1)}\left(\average{N^{2}\rho}+\average{N^{2}p}\right)\\
 & &-\average{N^{2}(v^{k}-N^{k})\tilde{\nabla}_{k}(\rho +p)}-\average{N^{3}(2v\dot{v}+v_{k}a^{k})(\rho+p)}-\average{N^{3}v^{2}\partial_{t}p}\nonumber\mbox{ .}
\end{eqnarray}
Finally, one can obtain an integrability condition for the system formed by Eqs. (\ref{eq:Averaged1}) and (\ref{eq:Averaged2}):
\begin{eqnarray}
 & &\partial_{t}{\cal Q}_{\CD}-\partial_{t}{\cal L}_{\CD}+\partial_{t}\average{\gamma^{2}N^{2}\CR}-2\Lambda\partial_{t}\average{N^{2}\gamma^{2}}+2H_{\CD}\left({\cal Q}_{\CD}-3{\cal L}_{\CD}+\average{\gamma^{2}N^{2}\CR}+2{\cal P_{\CD}}+2{\cal K}_{\CD}+2{\cal F}_{\CD}\right)\nonumber\\
\label{eq:Integrability}
 & &=16\pi G\left\{\partial_{t}\average{\gamma^{4}N^{2}\rho}+H_{\CD}\left[\average{N^{2}\rho\gamma^{3}\left(2\gamma+1+v^{2}\right)}+\average{N^{2}p\left(\gamma^{4}v^{2}+3\gamma(1+2\gamma^{2}v^{2})\right)}\right]+\partial_{t}\average{\gamma^{4}v^{2}N^{2}p}\right\}\mbox{ .}
\end{eqnarray}
Unfortunately, when $v^{k}$ is not zero, this last equation does not simplify because $\rho$ and $p$ appear with different space-time dependent coefficients in (\ref{eq:Averaged1}) and (\ref{eq:Averaged2}), so that one cannot make use of (\ref{eq:Averaged3}). Nevertheless, it is easy to see that when $v^{k}=0$ and $N^{k}=0$, the system (\ref{eq:Averaged1})-(\ref{eq:Averaged2})-(\ref{eq:Averaged3})-(\ref{eq:Integrability}) reduces to the well known results of \cite{Thomas:PerfFluid}, i.e. that the terms ${\cal L}_{\CD}$, ${\cal K}_{\CD}$ and ${\cal F}_{\CD}$ identically vanish. Another remark on this system: one can note that ${\cal Q}_{\CD}$ and ${\cal L}_{\CD}$ appear everywhere in this system in the combination ${\cal Q}_{\CD}-{\cal L}_{\CD}$, so that in a tilted coordinate system, it is more natural to define the kinematical backreaction as ${\cal Q}_{\CD}-{\cal L}_{\CD}$, ${\cal L}_{\CD}$ taking into account the pure velocity effects; nevertheless, we chose to stick to the conventions of \cite{Thomas:PerfFluid} so that the contact with the comoving observers can be done easily. 
First, if applied to the Newtonian gauge of perturbation theory, this formalism is expected to differ from previous analyses, in particular the one of \cite{Iain}, because of the additional terms ${\cal L}_{\CD}$, ${\cal K}_{\CD}$ and ${\cal F}_{\CD}$. Again, this comes from the fact that they defined the Hubble parameter with the expansion of the observers rather than with the expansion of the matter fluid. 
Second, the model proposed in \cite{Wiltshire} relies mainly on the fact that the lapse function is not constant throughout space-time due to the status of observers in virialized regions with respect to averaged cosmological observers identified to be void observers. But it is clear that if the void observers define the normal vector $n^{a}$, then the virialized observers must have a peculiar velocity with respect to them, so that the new terms also have to be estimated in this context if one wants to have a fully consistent framework.

\section{Conclusion and Outlook}
In this paper, we have generalized the averaging formalism of \cite{ThomasDust, Thomas:PerfFluid} to account for a possible tilt between the cosmological matter flow and the set of observers for which the (spatial) cosmological principle is supposed to hold. We have shown that, even if the general picture is not dramatically changed, the averaged system of equations governing the behavior of the effective model on large scales involves a lot of nontrivial new terms that describe the velocity effects (as well as the effect of the presence of a shift vector in the metric). Our formalism, based on the standard Riemannian averaging, is coordinate dependent, as pointed out in a recent paper \cite{Veneziano}. In this paper, the authors have shown that, in perturbation theory, the crucial point about the gauge dependence of the formalism lies in the change in the domain of averaging and they have explicitly built an alternative averaging procedure that is gauge independent. Therefore, in order to understand the physical meaning of the gauge dependence of our formalism, it will be employed in forthcoming works to study the backreaction from perturbations in the Newtonian gauge, and then to analyze the differences in the backreaction effect when one try and evaluate it in different perturbative gauges. A comparison of the results in the different gauges with the ones of the gauge independent procedure of \cite{Veneziano} also seems important.
Moreover, the formalism presented here will be at the heart of an investigation of the link between lightcone and spatial averagings, a link that is necessary to put observational constraints on averaged cosmological models. Finally, the author thinks that this will be useful in describing more precisely Wiltshire's idea about the difference between cosmological observers and matter observers in virialized regions \cite{Wiltshire}.

\section*{Acknowledgments}
This work is supported by a Claude Leon Foundation fellowship. The author is very grateful to K. Ananda, T. Buchert, C. Clarkson, P. Dunsby, G. Ellis,  A. Hamilton,  C. Hellaby and R. Maartens for useful discussions and comments during the preparation of this paper.

\end{document}